\documentclass[prd,amssymb,amsmath,amsfonts,aps,nofootinbib,notitlepage]{revtex4}
\pdfoutput=1
\usepackage{amsmath,amssymb}
\usepackage{graphicx}
\usepackage{enumerate}
\usepackage[colorlinks=true]{hyperref}
\usepackage{multirow} 
\usepackage{comment}
\includecomment{notes}
\specialcomment{notes}
{\begingroup}{\endgroup}
\excludecomment{notes}

\newcommand{\AddrUdeA}{
  {\it Instituto de F\'\i sica, Universidad de Antioquia,
    A.A.{\it{1226}}, Medell\'\i n, Colombia}}

\begin{document}
\title{Baryonic violation of $R$-parity from anomalous $U(1)_H$}
\author{Andres Florez}\email{florez@gfif.udea.edu.co}\affiliation{\AddrUdeA}
\author{Diego Restrepo}\email{restrepo@udea.edu.co}\affiliation{\AddrUdeA}
\author{Mauricio Velasquez}\email{mjvelas@gfif.udea.edu.co}\affiliation{\AddrUdeA}
\author{Oscar Zapata}\email{ozapata@fisica.udea.edu.co}\affiliation{\AddrUdeA}
\date{\today}   
\begin{abstract}
  Supersymmetric scenarios with $R$-parity conservation are becoming
  very constrained due to the lack of missing energy signals
  associated to heavy neutral particles, thus motivating scenarios
  with $R$-parity violation.
  In view of this, we consider a supersymmetric model with $R$-parity
  violation and extended by an anomalous horizontal $U(1)_{H}$
  symmetry%
. A self-consistent framework with baryon-number
  violation is achieved along with a proper supression for lepton
  number violating dimension-5 operators, so that the proton can be
  sufficiently stable. 
  With the introduction of right-handed neutrinos both Dirac and
  Majorana masses can be accommodated within this model. The
  implications for collider physics are discussed.
\end{abstract}
\maketitle

\section{Introduction} 
In contrast to the standard model (SM), its supersymmetric version
(SSM) does not have accidental lepton ($L$) and baryon-number ($B$)
symmetries, and this can lead to major phenomenological problems, like
fast proton decay.  
The standard solution to forbid all dangerous operators is the imposition of a
discrete symmetry, like $R$--parity, and only in this minimal version
(MSSM) the lightest supersymmetric particle (LSP), generally the
neutralino, is stable, providing a good dark matter candidate.
However, the recent results on searches for supersymmetry by 
CMS~\cite{:2012mfa} and ATLAS \cite{:2012rz}
experiments have raised the bound on scalar and gluino masses, when
they are approximately equal, to the order of 1.4 TeV for scenarios
such as the $R$-parity conserving constrained minimal supersymmetric
standard model (CMSSM). 
These searches are mainly based on missing transverse momentum carried
by the LSP.
A high mass scale for scalars and gluinos represents a potential chink
in the initial proposal of the SSM as a possible solution to the
hierarchy problem.

However, these mass limits can be avoided in alternative supersymmetric
models such as the $R$-parity violating SSM
\cite{Hall:1983id,Ross:1984yg,Barger:1989rk,Dreiner:1997uz,Allanach:2003eb,Barbier:2004ez},
where the LSP is usually assumed to be the gravitino which also
provides a good decaying dark matter candidate
\cite{Takayama:2000uz,Buchmuller:2007ui}.
The next to the lightest supersymmetric particle (NLSP) decays to standard
model particles, and thus the missing transverse momentum may be
considerably reduced
\cite{deCampos:2007bn,Brust:2011tb,Butterworth:2009qa,Allanach:2012vj,Brust:2012uf,Asano:2012gj,Curtin:2012rm,Graham:2012th}. 
In addition, if the involved couplings are small enough, the presence of
displaced vertices may reduce the efficiency of the standard searches
at LHC~\cite{deCampos:2007bn,Graham:2012th}.
In particular, $R$-parity breaking scenarios with operators that violate
$B$ lead to the most difficult signals to be searched at
hadron colliders
\cite{Butterworth:2009qa,Brust:2011tb,Allanach:2012vj,Brust:2012uf,Asano:2012gj,Curtin:2012rm,Han:2012cu}.
The \textit{ad hoc} choice of a discrete symmetry, like Lepton
parity, to forbid all the $L$ violating operators give rise to several issues.
First, the size of the $R$-parity breaking
couplings must be chosen precisely by hand in order to avoid
constraints from flavor physics observables and other precision
physics observables~\cite{Barbier:2004ez}. 
Second, dimension-5 $L$ violating operators are automatically
forbidden, and lepton-number violating neutrino mass terms cannot be
generated at the renormalizable and nonrenormalizable
level~\cite{Ibanez:1991pr,Dreiner:2005rd}.

Thus, it will be desirable to build a general
framework for supersymmetric models with baryonic violation of $R$-parity
rather than from {\it ad hoc} choices as it was done in
\cite{Tamvakis:1996np,Eyal:1999gq,Nikolidakis:2007fc,Luhn:2007gq,Csaki:2011ge,KerenZur:2012fr,FileviezPerez:2011pt,Ruderman:2012jd,Dreiner:2012ae,Krnjaic:2012aj,Csaki:2013we,Franceschini:2013ne}. 
This is the purpose of this work.

We address this issue by considering the SSM
extended with an anomalous horizontal $U(1)_{H}$ symmetry {\it \`a la}
Froggatt-Nielsen (FN)~\cite{Froggatt:1978nt}.  In this kind of models
the standard model particles and their superpartners do not carry a
$R$-parity quantum number and carry a horizontal charge
($H$-charge) instead. 
In addition, these kinds of models involve new heavy FN fields and, in
the simplest realizations, an electroweak singlet superfield $S$ of
$H$-charge~$-1$, called the flavon. 
For a recent discussion see~\cite{Sierra:2009zq}\footnote{%
In models where two flavons are used to explain the charged 
fermion hierarchy,  it is possible to obtain  Lepton parity as a 
remnant from an horizontal symmetry \cite{Eyal:1999gq}}.
In the case of supersymmetric models based on an anomalous $U(1)_H$
flavor symmetry with a single flavon, the quark masses, the quark
mixing angles, the charged lepton masses, and the conditions of
anomaly cancellation constrain the possible $H$-charge assignments.
Since the number of constraints is always smaller than the number of
$H$-charges some of them are necessarily unconstrained and apart from
theoretical upper bounds on their values they can be regarded as
free parameters that should  be determined by additional
phenomenological input (for a review see \cite{Dreiner:2003hw}). 
This freedom can be used to set the order of magnitude of the $R$-parity
violating couplings.
Along these lines, consistent models have been built
in which neutrino oscillation data can be explained
\cite{Dreiner:2003hw,Choi:1998wc,Mira:2000gg,Dreiner:2003yr,Dreiner:2006xw,Dreiner:2007vp}. 
Also, by using the reported anomalies in cosmic-ray electron/positron
fluxes, a consistent model with tiny $R$-parity breaking couplings was
built with decaying leptophilic-neutralino dark
matter~\cite{Sierra:2009zq}.

We adopt a new approach here by assuming a set of $H$-charges which
give rise to a self-consistent model of $R$-parity breaking and
baryon-number violation.
As a consequence of our $H$-charge assignments it is not
possible to generate a Majorana mass term for left-handed neutrinos. 
However a neutrino Dirac matrix can be built after the introduction of
right-handed neutrinos with proper $H$-charges.
We also show that by adding a second flavon field with fractional
charge, it is possible to build a Majorana neutrino mass matrix.
In both cases an anarchical matrix
\cite{Hall:1999sn,Haba:2000be,deGouvea:2003xe,deGouvea:2012ac} is
obtained, which is supported by the recent results of a large value
for $\theta_{13}$ \cite{Abe:2011sj,Abe:2011fz,An:2012eh,Ahn:2012nd}.

As a consequence of $H$-charge assignments, the $\lambda''_{323}$
coupling dominates over the other couplings, and the third generation
quarks are expected to be present at the final states of LSP decays.  Moreover,
the horizontal symmetry predicts a precise hierarchy of $B$ violating
couplings which can be translated into relations between different
branching ratios, that could be measured at $e^+\ e^-$ colliders.

In the next Section the required conditions to obtain one
$R$-parity breaking SSM with $B$ violation are shown, taking into account also
dimension-5 operators.
The generation of neutrino masses by introducing right-handed
neutrinos is discussed in 
section~\ref{sec:RHN}. In Section~\ref{sec:implications}
the consequences for collider physics are mentioned, and then Section \ref{sec:conlusions} ends with the conclusions. 
In the appendices the horizontal charges of the
dimension-4 and dimension-5 R-parity breaking operators are detailed.



\section{Horizontal model with Baryon number violation}
\label{sec:BNVmodel}
To solve the charged fermion mass hierarchy in the SSM, it is used to
invoke the FN mechanism \cite{Froggatt:1978nt}. 
In the simplest scenario, the $U(1)_H$ symmetry is spontaneously
broken at one scale close to Planck mass, $M_P$, by the vacuum expectation
value of a SM singlet scalar, the flavon field $S$, with $H$-charge~$-1$, which allows to define the expansion parameter $\theta=\langle S
\rangle/M_P\approx0.22$ \cite{Dreiner:2003yr,Irges:1998ax}. 
The fermion masses and mixings are determined by factors of the type
$\theta^n$, where $n$ is fixed by the horizontal charges of the fields
involved.
In supersymmetric scenarios, the order of magnitude of the $R$-parity
violating couplings can also be fixed by the FN mechanism
\cite{Dreiner:2003hw,Choi:1998wc,Mira:2000gg,Dreiner:2003yr,Choi:1996se,Joshipura:2000sn,Binetruy:1996xk,Ellis:1998rj,BenHamo:1994bq}.

The most general renormalizable superpotential respecting the gauge
invariance of the standard model is given by
\cite{Allanach:2003eb,Barbier:2004ez}
\begin{align}
  \label{eq:W}
  W =&  
  h^u_{ij}\widehat{H}_u\widehat{Q}_i\widehat{u}_j + h^d_{ij}\widehat{L}_0\widehat{Q}_i\widehat{d}_j+h^l_{ij} \widehat{L}_0\widehat{L}_i\widehat{l}_j\nonumber\\
  &+\mu_\alpha\widehat{L}_\alpha\widehat{H}_u+\tfrac{1}{2}\lambda_{ijk}\widehat{L}_i\widehat{L}_j\widehat{l}_k +  \lambda'_{i j k}\widehat{L}_i\widehat{Q}_j\widehat{d}_k + 
  +\tfrac{1}{2}\lambda''_{ijk}\widehat{u}_i\widehat{d}_j\widehat{d}_k
  \,,
\end{align}
where $i, j, k=1,2,3$; $\alpha=0,\ldots,3$, and the down-type Higgs
superfield $\widehat{H}_d$ is denoted by $\widehat L_0$.  
Lepton-number is explicitly broken by the bilinear couplings $\mu_i$
and trilinear couplings $\lambda_{ijk}$ and $\lambda_{ijk}'$, whereas
the couplings $\lambda_{ijk}''$ are responsible for the $B$ violation.
The factor of $1/2$ are due to the antisymmetry of the corresponding
operators~\cite{Barbier:2004ez}.
The $H$-charges for the fields determines whether or not a particular term in
the second line of eq.~\eqref{eq:W} can be present in the
superpotential.

Before proceeding we will fix our notation: following Ref.
\cite{Mira:2000gg} we will denote a field and its $H$--charge with the
same symbol, i.e.  $H(f_i)=f_i$, $H$--charge differences as
$H(f_i-f_j)=f_{ij}$ \cite{Dudas:1995yu}, and bilinear $H$--charges as
$n_\alpha = L_\alpha + H_u$.
In what follows we will constrain the $H$-charges to satisfy the
condition $|H(f_i)| < 10$ which leads to a consistent prediction of
the size of the suppression factor $\theta$ in the context of string
theories~\cite{Dreiner:2003hw,Choi:1996se}.
To properly account for the hierarchy of charged fermions with a
single flavon, the symmetry $U(1)_H$ need to be anomalous.
With three theoretical restrictions coming from anomaly cancellation 
through the Green-Schwarz mechanism~\cite{Green:1984sg},
eight phenomenology conditions from mass ratios and mixings of the
charged fermionic sector, and two more conditions corresponding to the
absolute value of the third generation fermion masses, we obtain a set
of 13 conditions.
Hence, 13 out of 17 $H$-charges are constrained and can be expressed
in terms of the remaining four charges that have chosen to be the lepton
number violating bilinear $H$-charges $n_{i}$, and $x$
\cite{Mira:2000gg}, where $x=L_0+L_3+e_3=L_0+Q_3+d_3$ takes integer
values from 0 to 3 in order to obtain the allowed range for
$\tan\beta=\theta^{x-3}$. 
With all this restrictions, there is only a possible set of charge
differences which is displayed in Table~\ref{tab:chargediff}. This
self-consistent solution includes the Guidice-Masiero mechanism to
solve the $\mu$ problem because $n_0=-1$, and therefore the $\mu$ term
is absent from the superpotential~\cite{Mira:2000gg}.

\begin{table}[t]
  \centering
  \begin{tabular}{cccccccc}\hline
    $Q_{13}$&$Q_{23}$&$d_{13}$&$d_{23}$&$u_{13}$&$u_{23}$&$\mathcal{L}_{13}$&$\mathcal{L}_{23}$\\\hline
    $3$&$2$&$1$&$0$&$5$&$2$&$5$&$2$\\\hline    
  \end{tabular}
  \caption{Standard model fields $H$--charges differences with $n_0=-1$ (from ref. \cite{Mira:2000gg}). Here $\mathcal{L}_{i3}=L_{i3}+l_{i3}$}
  \label{tab:chargediff}
\end{table}

The $H$-charges of the $R$-parity breaking
couplings can be written as 
\begin{align}
\label{eq:Hrpv}
  H(\lambda''_{ijk})=&\tfrac{1}{3}\mathcal{N}+\left[x+\mathcal{I}''(i,j,k)\right]&(j<&k)\nonumber\\
  H(\lambda'_{ijk})=&n_i+\left[x+\mathcal{I}'(i,j,k)\right]                      &&     \nonumber\\
  H(\lambda_{ijk})=&n_i+n_j-n_k+n_0+\left[x+\mathcal{I}(i,j,k)\right]           &(i<&j)\nonumber\\
  =&
  \begin{cases}
    n_{i(\text{or $j$})}+n_0+\left[x+\mathcal{I}(i,j,k)\right]& \text{if $i= k$ (or $j= k$)}\\
    \mathcal{N}-2n_k+\left[x+\mathcal{I}(i,j,k)\right]& \text{if $i\ne k$ and $j\ne k$}\\
  \end{cases},&&
\end{align}
where
\begin{align}
  \mathcal{N}=\sum_{\alpha=0}^3 n_\alpha=n_0+n_1+n_2+n_3=n_1+n_2+n_3-1\,,
\end{align}
is the sum of the bilinear $H$-charges.
The terms inside the brackets in eq.~\eqref{eq:Hrpv}, are the integer
part of the corresponding $H$-charges, with the $\mathcal{I}$'s being
functions of the coupling indices returning \emph{integer} values.
They are given explicitly in eq.~\eqref{eq:Is}.

From eq.~\eqref{eq:Hrpv} is straightforward to see the possible
scenarios in the context of an anomalous horizontal Abelian symmetry
with a single flavon, reviewed in the introduction.
The MSSM is obtained when $\mathcal{N}/3$, each individual $n_i$  
and $\mathcal{N}-2n_k$ are fractional~\cite{Dreiner:2003yr,Dreiner:2007vp}.
Bilinear $R$-parity violation\footnote{See for example \cite{Diaz:2003as} and references therein} is obtained when $\mathcal{N}/3$
is fractional and each $n_i$ is a negative integer ~\cite{Mira:2000gg,Dreiner:2006xw}.
Another self-consistent $R$-parity breaking model with $L$ violation can
be obtained if $\mathcal{N}/3$ and  each individual $n_i$ are fractional,
but some of the  $\mathcal{N}-2n_k$ are integer. In such a case the
decays of the LSP are leptophilic~\cite{Sierra:2009zq}.

In this work we want to explore the last self-consistent possibility,
consisting in the $R$-parity breaking model with $B$-violation.  It is
clear from eq.~\eqref{eq:Hrpv} that if $\mathcal{N}$ is integer and
multiple of 3, and each $n_i$ is fractional but not half-integer,
then only the 9 $\lambda_{ijk}''$ are generated.  The specific horizontal charges are:
\begin{align}
\label{eq:Hlpp}
 H \begin{pmatrix}
    \lambda_{112}''&\lambda_{212}''&\lambda_{312}''\\
    \lambda_{113}''&\lambda_{213}''&\lambda_{313}''\\
    \lambda_{123}''&\lambda_{223}''&\lambda_{323}''\\
  \end{pmatrix}=
 \begin{pmatrix}
   6 & 3 & 1\\
   6 & 3 & 1\\
    5 & 2 & 0\\
  \end{pmatrix} +n_{\lambda''}\mathbf{1_{3}},
\end{align}
where  $\mathbf{1_3}$ is a $3\times 3$ matrix filled with ones, and  $n_{\lambda''}$  is defined by
\begin{align}\label{nlpp}
  n_{\lambda''}=x+ \frac{1}{3}\mathcal{N}\,.
\end{align}

For positive $n_{\lambda''}$ values, the third generation couplings dominate with fixed ratios between them:
\begin{align}
\label{eq:Hierlpp}
 \begin{pmatrix}
    \lambda_{112}''&\lambda_{212}''&\lambda_{312}''\\
    \lambda_{113}''&\lambda_{213}''&\lambda_{313}''\\
    \lambda_{123}''&\lambda_{223}''&\lambda_{323}''\\
  \end{pmatrix}\approx&
 \theta^{n_{\lambda''}}\begin{pmatrix}
   \theta^6 & \theta^3 & \theta\\
   \theta^6 & \theta^3 & \theta\\
   \theta^5 & \theta^2 & 1\\
  \end{pmatrix} & n_{\lambda''}\ge 0\,.
\end{align}
For negative values some of the couplings start to be forbidden in the superpotential by
holomorphy, and for $n_{\lambda''}<-6$ all of them must be generated
from the K\"ahler potential with additional Planck mass suppression,
so that the LSP may be a decaying dark matter candidate as in the
case of $L$ violation studied in \cite{Sierra:2009zq}.
We will not pursue this possibility in this work because in that case
the phenomenology at colliders should be the same than in the MSSM.

Below the allowed range for $n_{\lambda''}$ and their
consequences at present and future colliders will be checked.

\subsection{Constraints from $\Delta B\neq0$ processes}

\begin{figure}
  \centering
  \includegraphics[scale=0.6]{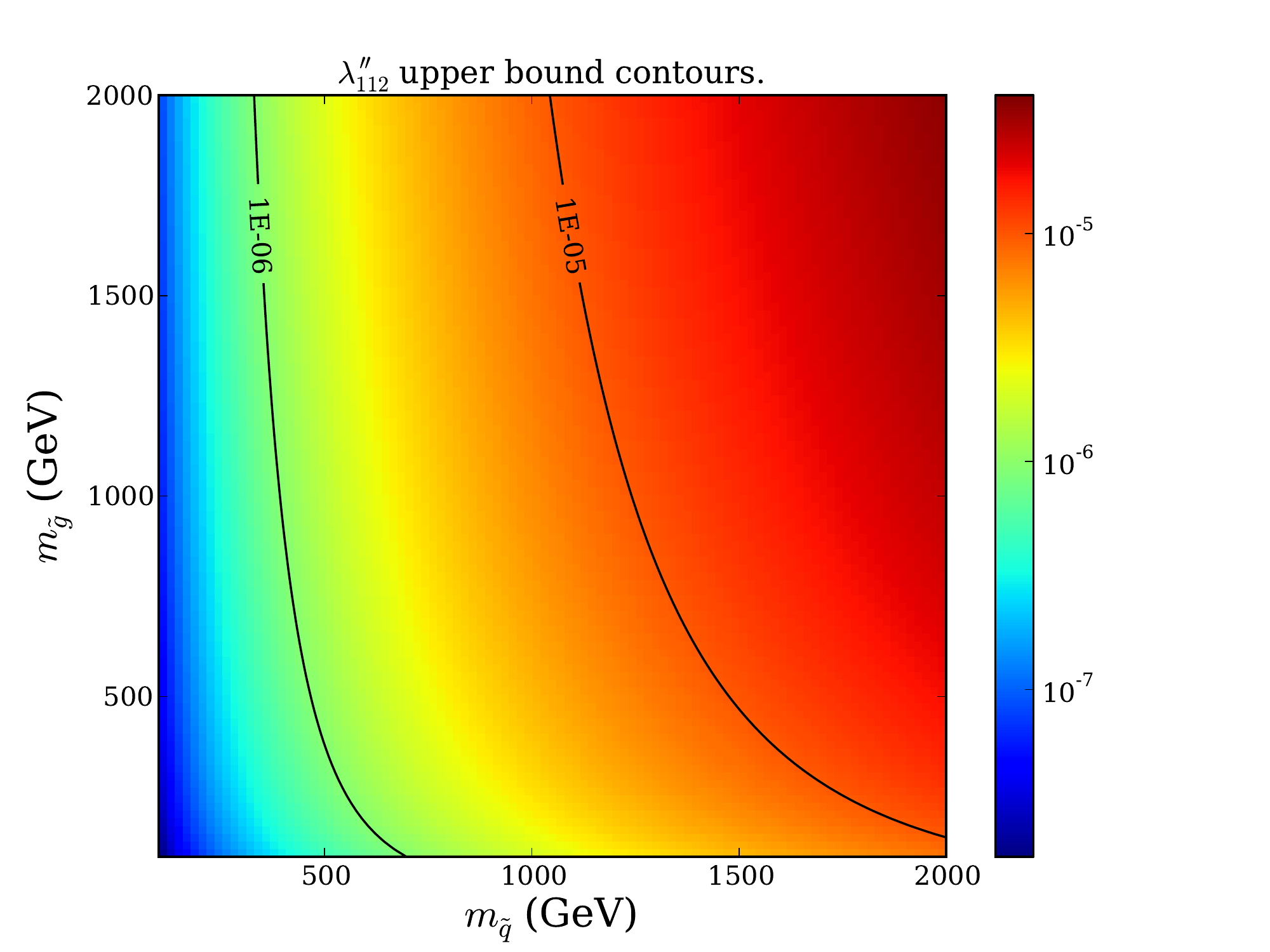}
  \caption{$\lambda_{112}''$ constraint as a function of squark and gluinos mass, for $\tilde{\Lambda}=75\ \text{GeV}$ and $\tau_{NN\to KK}=1.7\times10^{32}\, \text{years}$}
  \label{fig:constraints_contour}
\end{figure}

Several experimental constraints are found on $B$ violating
couplings both for individual and quadratic product of couplings \cite{Barbier:2004ez}. For individual couplings, the stronger constraints are for  $\lambda_{11k}$. Because in our model the predicted order of magnitude for the coupling $\lambda_{113}''$ is the same that for $\lambda_{112}''$, the most restrictive constraint is the obtained for the later, and comes from 
the dinucleon $NN\to KK$ width, which accordingly to  \cite{Goity:1994dq,Csaki:2011ge} is
\begin{align}
  \Gamma\sim \rho_N\frac{128\pi\alpha_s^2|\lambda_{112}|^4(\tilde{\Lambda})^{10}}{m_{N}^2m_{\tilde{g}}^2m_{\tilde{q}}^8}\,,
\end{align}
where $\rho_N\approx 0.25\ \text{fm}^{-3}$ is the nucleon density,
$m_N\approx m_p$ is the nucleon mass, and $\alpha_s\approx0.12$ is the
strong coupling.
Note that this kind of matter instability requires only $B$ violation, and is suppressed by the tenth power of  $\tilde{\Lambda}$, which parameterizes the hadron and nuclear effects. 
For this quantity,  order of magnitude variation is expected around of the $\Lambda_{\text{QCD}}$ scale of $200\ \text{MeV}$. 
However, $\tilde{\Lambda}$ is roughly expected to be smaller than $\Lambda_{\text{QCD}}$ because the repulsion effects inside the nucleus~\cite{Csaki:2011ge}.  
From general experimental searches of matter instability~\cite{Berger:1991fa}, lower bounds similar to the proton life time should be used  for this specific dinucleon channel~\cite{Goity:1994dq}, and therefore additional suppression from $\lambda_{112}$ could be required. In fact, the first lower bound on dinucleon decay to kaons have been recently obtained from Super-Kamiokande data~\cite{Litos:2010}
\begin{align*}
  \tau_{NN\to KK} =\frac{1}{\Gamma}>1.7\times10^{32}\ \text{years}\,.
\end{align*}
From this value, we can obtain  a constraint for the $B$ violating coupling 
\begin{align}
  |\lambda_{112}|\lesssim3.2\times 10^{-7}
\left(\frac{1.7\times10^{32}\, \text{years}}{\tau_{NN\to KK}}\right)^{1/4}
\left(\frac{m_{\tilde{g}}}{300\, \text{GeV}}\right)^{1/2}
\left(\frac{m_{\tilde{q}}}{300\, \text{GeV}}\right)^{2}
\left(\frac{75\ \text{MeV}}{\tilde{\Lambda}}\right)^{5/2}\,,
\end{align}
where a conservative value for $\tilde{\Lambda}$, as in \cite{Barbier:2004ez}, has been used. Large values of $\tilde\Lambda$ give arise to even smaller upper bounds for $|\lambda_{112}|$.  In figure~\ref{fig:constraints_contour}, it is illustrated the effect of varying gluino and squark masses.  We can see that the constraint still holds strong for large values of the relevant supersymmetric masses, especially for low-mass gluinos. 

\begin{figure}
  \centering
  \includegraphics[scale=0.7]{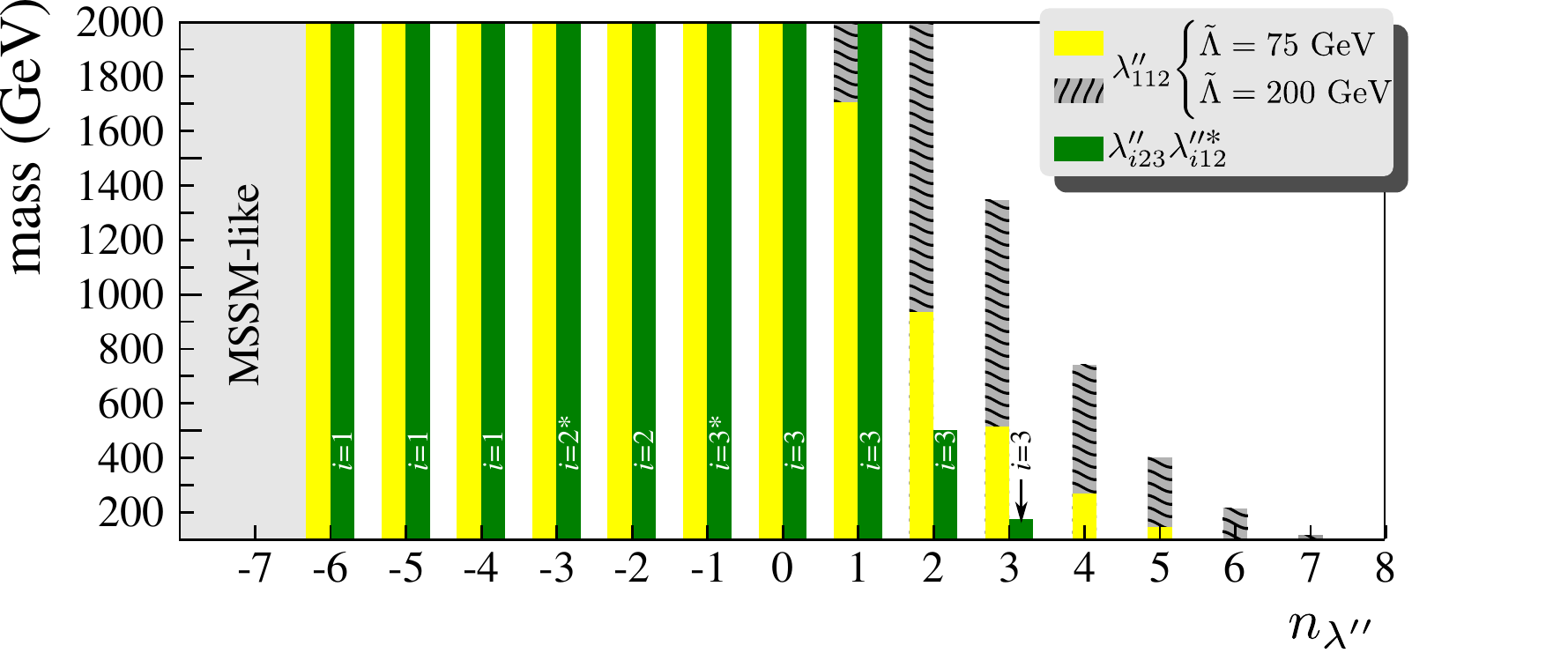}
  \caption{The yellow bands (green bands) display the excluded range for 
    $\tilde{m}$ ($\tilde{u}_{iR}$), as a function of the possible $n_{\lambda''}$ 
    solutions from the constraints in 
    $|\lambda_{112}''|$ ($|\lambda_{i23}''\lambda_{i12}^{\prime\prime *}|$). 
    The gray-dashed bands show the effect of increasing
    $\tilde{\Lambda}$ in the $|\lambda_{112}''|$ constraint. 
    The affected $\tilde{u}_{iR}$ is indicated for each value of
    $n_{\lambda''}$. 
    For $n_{\lambda''}<-6$, the phenomenology at
    colliders is expected to the same as in the MSSM. }
  \label{fig:constraints}
\end{figure}

For $\tilde{m}=m_{\tilde{g}}=m_{\tilde{q}}$, we can obtain the lower bound
\begin{align}
  \label{eq:tildem}
 \tilde{m}  \gtrsim& (279\ \text{GeV})\theta^{(-8+2n_{\lambda''})/5}
   \left(\frac{\tau_{NN\to KK}}{1.7\times10^{32}\, \text{years}}\right)^{1/10}
   \left(\frac{\tilde{\Lambda}}{75\ \text{MeV}}\right),&n_{\lambda''}\ge& -6\,.
\end{align}
The excluded supersymmetric masses as function of $n_{\lambda''}$, are illustrated with the yellow (light-gray) bands in figure~\ref{fig:constraints}. The important restrictions appear for negative powers of $\theta$ in eq.~\eqref{eq:tildem}, corresponding to $n_{\lambda''}\le 4$. If $\tilde\Lambda$ is increased until $150\ \text{GeV}$,  stronger restrictions are obtained, as illustrated in the dashed bands of figure~\ref{fig:constraints}. We can see that for the full range of equal gluino and squark  masses displayed in figure~\ref{fig:constraints}, the constraint is strong enough to forbid all the negative solutions of $n_{\lambda''}$, and also some of the positive solutions depending of the chosen $\tilde{\Lambda}$ value.

It is also possible to exclude the negative solutions if we use the available quadratic coupling product bounds. For our model the most important constraint is obtained from the penguin decays $B\to \phi\pi$~\cite{BarShalom:2002sv,Barbier:2004ez}. Updating the limit with the last result from BaBar~\cite{Aubert:2006nn}\footnote{The limit from Belle is $\operatorname{Br}(B^+\to \phi\pi^+)<3.3\times 10^{-7}$~\cite{Kim:2012gt}.} to $\operatorname{Br}(B^+\to \phi\pi^+)<2.4\times10^{-7}$, we obtain from fig. 3 of~\cite{BarShalom:2002sv}
\begin{align}
  |\lambda_{i23}''\lambda_{i12}^{\prime\prime *}|<2\times 10^{-5}\left(
    \frac{m_{\tilde{u}_{i R}}}{100\text{ GeV}}
  \right)^2\,.
\end{align}
The excluded right-handed up-squark masses are shown in the green (dark gray) bands
of figure~\ref{fig:constraints}, with the specific generation of up-squark
labeled inside the band.
The solutions with the additional ``*''-label, have the  quoted $\lambda_{i23}''$ coupling  absent
from the superpotential. However it is regenerated at order $\theta$
through a K\"ahler rotation~\cite{Choi:1996se} from the dominant
coupling still present in the superpotential.
As a result, again the negative solutions are excluded for the full range of squark
masses displayed in the figure. 
Moreover, the first two positive solutions are also excluded. 
In the figure, the gray region for $n_{\lambda''}\le-7$ is also shown.
In this case, the holomorphy of the superpotential forbids all the
$\lambda''$ terms and although they will be generated after $U(1)_H$
symmetry breaking via the K\"ahler potential \cite{Giudice:1988yz},
these terms are suppressed by the additional factor
$m_{3/2}/M_P$~\cite{Sierra:2009zq}.
Therefore the LSP is very long-lived and the phenomenology at
colliders is expected to be the same than in the MSSM.

Therefore, by demanding a $B$ violating model and imposing the
constraints on the $R$-parity breaking couplings, only positive
solutions for $n_{\lambda''}$ remains allowed giving rise to a clear
hierarchy between $\lambda''$ couplings, which have a direct impact on
the phenomenology of the LSP. 
The dominant coupling turns out to be $\lambda''_{323}$, a feature
shared with \cite{Csaki:2011ge,KerenZur:2012fr}.

\subsection{Dimension-5 operators and proton decay}
So far the $U(1)_H$ symmetry has been used to forbid dimension-4 lepton number
violating couplings, in order to keep proton decay to a safe limit.
However,  proton decay mediated by $\lambda''$ couplings alone, can occurs in
scenarios with a gravitino lighter than proton \cite{Choi:1996nk},
leading to strong bounds on these couplings. Thus, by ensuring gravitino
masses greater than 1 GeV in these scenarios there will be no
contribution to the proton decay coming from gravitino, 
which being the LSP can be also a dark matter canditate
\cite{Takayama:2000uz,Buchmuller:2007ui,Csaki:2011ge,Lola:2008bk,Lola:2007rw}.

On the other hand, there are also dimension-5 lepton or$/$and baryon
number violating couplings, which can induce proton decay.  Hence, it
is also necessary to check if these terms also are banned or enough
suppressed.
   
The non-renormalizable dimension-5 operators in the superpotential
$W_{5D}$ and K\"ahler potential $V_{5D}$ are given by
\cite{Ibanez:1991pr,Ibanez:1991hv,Allanach:2003eb,Barbier:2004ez}
\begin{align}\label{WD5}\nonumber
W_{D5}&=\frac{(\kappa_1)_{ijkl}}{M_P}\widehat{Q}_i\widehat{Q}_j\widehat{Q}_k\widehat{L}_l+\frac{(\kappa_2)_{ijkl}}{M_P}\widehat{u}_i\widehat{u}_j\widehat{d}_k\widehat{e}_l+\frac{(\kappa_3)_{ijk}}{M_P}\widehat{Q}_i\widehat{Q}_j\widehat{Q}_k\widehat{H}_d\\
&+\frac{(\kappa_4)_{ijk}}{M_P}\widehat{Q}_i\widehat{H}_d\widehat{u}_j\widehat{e}_k+\frac{(\kappa_5)_{ij}}{M_P}\widehat{L}_i\widehat{H}_u\widehat{L}_j\widehat{H}_u+\frac{(\kappa_6)_{i}}{M_P}\widehat{L}_i\widehat{H}_u\widehat{H}_d\widehat{H}_d,\\
V_{5D}&=\frac{(\kappa_7)_{ijk}}{M_P}\widehat{u}_i\widehat{d}^{*}_j\widehat{e}_k+\frac{(\kappa_8)_{i}}{M_P}\widehat{H}^{*}_u\widehat{H}_d\widehat{e}_i+\frac{(\kappa_9)_{ijk}}{M_P}\widehat{Q}_i\widehat{L}^{*}_j\widehat{u}_k+\frac{(\kappa_{10})_{ijk}}{M_P}\widehat{Q}_i\widehat{Q}_j\widehat{d}^{*}_k.
\end{align}

A review of the effect of this operators in the destabilization of the
proton is given in~\cite{Dreiner:2012ae}. 
In the present case of $B$ violation, we would guarantee a
sufficiently stable proton if the $B$ and $L$ violating  operators with couplings $\kappa_{1,2}$, 
and the $L$ violating operators with coupling $\kappa_{4,7,8,9}$ are forbidden\footnote{The constraints on the operator with coupling $\kappa_6$ are mild~\cite{Dreiner:2012ae}}.
The operator with coupling $\kappa_5$, $LH_uLH_u$, is not constrained
by proton decays because it violates lepton number by two units.

The horizontal charges for all the dimension-5 operators are given
in the Appendix \ref{app:D5}.
Given the fractional values ​​needed for $n_i$ in order to get rid of
the dimension-4 $L$ violating operators in eq.~\eqref{eq:W}, it turns
out that all dimension-5 $L$ violating operators are also
automatically forbidden by the $U(1)_H$ symmetry (see
Eqs. (\ref{eq:D5BLNV1}), (\ref{eq:D5BLNV2}) and (\ref{eq:D5LNV})).
At this stage the $U(1)_H$ symmetry plays the same role that a lepton
parity discrete symmetry \cite{Allanach:2003eb,Ibanez:1991pr,Ibanez:1991hv,Dreiner:2005rd}.



\section{Generation of neutrino masses}
\label{sec:RHN}
Although it is not required that $LH_uLH_u$ operator be forbidden
by $U(1)_H$ symmetry to ensure proton stability,
it is unavoidable prohibited because the
bilinears charges $n_i$ are not half-integers.  
Thus, the Majorana mass terms $\nu_L\nu_L$ are automatically forbidden.  
The same happens with Lepton parity symmetry, and also within the more general
approach of gauge discrete symmetries~\cite{Ibanez:1991pr,Ibanez:1991hv,Dreiner:2005rd},
where the solutions which allows the $UDD$ operator
automatically forbids Majorana neutrinos.
The proposed solution in this kind of frameworks is just to introduce
right-handed neutrinos $N$ with  their Majorana mass terms $NN$
forbidden, while keeping the Yukawa operators containing left and
right-handed neutrino still allowed, generating in this way Dirac neutrino
mass matrices~\cite{Luhn:2007gq}. 
When these ideas are applied to our case of horizontal symmetries, it is also necessary to explain the smallness of the neutrino
Yukawa couplings.
The introduction of three right-handed neutrinos $N_{i}$ $(i=1,2,3)$ allow us to give 
Dirac masses to neutrinos by assigning fractional and not half-integer
$H$-charges to $N_{i}$, such that the $NN$ terms remains forbidden.

Let us paramaterize the bilinear $H$-charges as $n_2=n_1+\alpha$,
$n_3=n_1+\beta$, and for right-handed neutrinos: $N_2=N_1+\epsilon$
and $N_3=N_1+\rho$. The neutrino Dirac mass matrix reads
\begin{align}
 M_{\nu}\sim 
v_u\theta^{\beta+\rho+n_1+N_1}
\left(
\begin{array}{ccc}
 \theta^{-\beta-\rho } & \theta^{\epsilon -\beta-\rho} & \theta^{-\beta} \\
 \theta^{\alpha -\beta-\rho } & \theta^{\alpha +\epsilon-\beta -\rho} & \theta^{\alpha-\beta} \\
 \theta^{-\rho} & \theta^{\epsilon-\rho } & 1
\end{array}
\right),
\end{align}
where  $v_{u}$ is the vacuum expectation value developed by up-type Higgs field. 
From Eq. (\ref{nlpp}) we obtain $n_1=\frac{1}{3}\left(1-\alpha -\beta
  +3 n_{\lambda ''}-3 x\right)$. Motivated by the recent results of a
large value for $\theta_{13}$
\cite{Abe:2011sj,Abe:2011fz,An:2012eh,Ahn:2012nd}, which support those
models based on a anarchical neutrino mass matrix
\cite{Hall:1999sn,Haba:2000be,deGouvea:2003xe,deGouvea:2012ac}, it is
convenient to choose $\alpha=\beta=\epsilon=\rho=0$ and
$\beta+\rho+n_1+N_1=n_{Y_\nu}$ with $n_{Y_\nu}$ being integer and $n_{Y_\nu}\ge 16$ 
in order to generate a neutrino Yukawa couplings 
$Y_\nu\lesssim10^{-11}$. 
It is worth stressing that since $n_1$ cannot be integer, the
$\mu\tau$ anarchical texture with $\alpha=\beta=-1$
\cite{Altarelli:2002sg,Dreiner:2003yr,Buchmuller:2011tm,Altarelli:2012ia}
is not allowed.
However, other textures can be accommodated in our model 
\cite{Altarelli:2012ia} such as pseudo $\mu\tau$-anarchy
($\alpha=\beta=\epsilon=\rho=-2$) and the hierarchical texture
($\alpha=\epsilon=-1,\, \beta=\rho=-2$).
An immediate consequence of the anarchy assumption is that the
bilinear charges are equal and are set to $n_i=n_{\lambda
  ''}-x+\frac{1}{3}$, being clearly non-integer numbers.
The $H$-charges which allow to obtain a self-consistent framework with
the requirements mentioned above are shown in table \ref{tab:Hc3}. 
It is remarkable that when explaining the neutrino Yukawa couplings
$Y_\nu$, a lower bound on $n_{\lambda''}\ge 6$ emerges, which leads
to deep implications on the phenomenology of the model (see next
Section).

\begin{table}[t]
  \centering
  \begin{tabular}{ccccccccc}\hline
    $x$            & 0 & 0 & 0 & 0 & 1 & 0 & 1 & 2 \\\hline 
    $n_{\lambda''}$ & 6 & 7 & 8 & 9 & 9 & 10 & 10& 10\\
    $n_i$ & 19/3 & 22/3 & $25/3$ & $28/3$ & $25/3$ & $31/3$ & $28/3$&$25/3$  \\
    $N_i$ & $29/3$ & $26/3$ & $29/3$ & $26/3$ & $29/3$ &$23/3$ & $26/3$&29/3   \\\hline 
      \end{tabular}
      \caption{Some sets of $H$-charges allowing a  self-consistent framework of 
        $R$-parity breaking with $B$ violation and Dirac neutrinos.}
  \label{tab:Hc3}
\end{table}
\subsection{Majorana neutrinos}
It is worth mentioning that it is also possible to have Majorana
neutrinos if in addition to the right-handed neutrinos we include in the
model a second flavon\footnote{For a model with several flavons see
  \cite{Jack:2003pb}}, $\psi$, with fractional\footnote{A scenario
  with Majorana neutrinos and non-anomalous $U(1)_H$ symmetry, which
  is spontaneously broken by two flavons with opposite $H$-charges +1
  and -1 was obtained in \cite{Eyal:1999gq}.} $H$-charge and with a
vacuum expectation value approximately equal to $\theta$.
The horizontal charges of these superfields are fixed by new invariant
diagrams coming from Dirac and Majorana mass terms.
 
In this way, the $H$-charge of $\psi$ must be such that it does not get
coupled to $L$ violating operators. Therefore, the respective total
$H$-charge of the full $L$ violating operator would be either
fractional and therefore forbidden, or negative and sufficiently
suppressed.

The introduction of an additional flavon field could spoil the proton
stability since $H$-invariant terms can be obtained by coupling a
large number of $\psi$ flavons to dangerous operators.
Therefore  it is  mandatory  to ensure  that  $L$ violating  bilinear,
dimension-4  and  dimension-5   operators  are  generated  through  GM
mechanism or have a large Froggatt-Nielsen suppression. 
The $H$-charges which allow to obtain Majorana neutrinos with
the requirements mentioned above, are shown in table \ref{tab:Hc}.  
To illustrate this point, let's consider the first solution given in table \ref{tab:Hc}. 
For that set of $H$-charges, we have found that the minimum suppression that is achieved 
for dimension-4 and 5 operators is 
$\widehat{L}_1\widehat{Q}_1\widehat{D}_1: m_{3/2}\theta^{21}/M_P$ and 
$\widehat{u}_1\widehat{u}_2\widehat{d}_1\widehat{e}_1: m_{3/2}\theta^2/M_P^2$, 
which is enough to satisfy the constraints coming from proton decay.

\begin{table}[t]
  \centering
  \begin{tabular}{ccccccccc}\hline
    $x$            & 1 & 1 & 1 & 2 & 2 & 2 & 3 & 3 \\\hline 
    $n_{\lambda''}$ & 5 & 6 & 7 & 6 & 7 & 8 & 8 & 9 \\
    $n_i$ & $13/3$ & $16/3$ & $19/3$ & $13/3$ & $16/3$ & $19/3$ & $16/3$ & $19/3$ \\
    $\psi$ & $-47/6$ & $-53/6$ & $-59/6$ &$-47/6$ & $-53/6$ & $-59/6$ &$-53/6$ & $-59/6$  \\\hline 
      \end{tabular}
      \caption{Sets of $H$-charges which allow having Majorana neutrinos with $H(N_i)=7/2$. 
        For this scenario there is no lower bound on $n_{\lambda''}$.}
  \label{tab:Hc}
\end{table}

Henceforth, we will combine the solutions allowed by the
experimental constraints on $R$-parity breaking couplings discussed in
section~\ref{sec:BNVmodel}, with the restrictions to obtain Dirac
neutrinos, and therefore we will only consider solutions with
$n_{\lambda''}\ge 6$.


\section{Implications on collider searches}
\label{sec:implications}
From a collider physics point of view, there are two main differences
between the models with and without $R$-parity conservation.
When $R$-parity conservation is assumed, the production
of supersymmetric particles is in pairs, and the LSP is stable leading
to missing energy signatures in the detectors.
On the other hand, $R$-parity violation allows for the single
production of supersymmetric particles and the decay of the LSP
involving jets or/and leptons.
The $R$-parity breaking and $B$ violating operators, induce LSP decay
directly or indirectly to quarks, including the top if LSP is
sufficiently massive\footnote{If a supersymmetric partner of some SM
  particle is the NLSP with the gravitino as the LSP, our
  phenomenological results would not change.}.
Given that the LSP is no longer stable due to R parity violation, in
principle the LSP can be any supersymmetric particle
\cite{Dreiner:1997uz, Barbier:2004ez,Dreiner:2008ca}.
For recent phenomenological studies in supersymmetric scenarios with
$R$-breaking through $B$ violating, see, {\it e.g.},
\cite{Butterworth:2009qa,Allanach:2012vj,Brust:2012uf,Asano:2012gj,Curtin:2012rm,Evans:2012bf,Berger:2012mm,Dreiner:2008ca,Carpenter:2008sy,Dreiner:2012wm,Franceschini:2012za,Bomark:2011ye,Lola:2008bk,Choudhury:2011ve,Desai:2010sq,Kilic:2011sr,Carpenter:2007zz,Kaplan:2007ap,Berger:2013sir}
and in particular \cite{Csaki:2011ge,KerenZur:2012fr}.

The phenomenology of the model at LHC is basically the same studied in
the SSM with Minimal Flavor Violation (MFV) \cite{Csaki:2011ge} and Partial Compositeness \cite{KerenZur:2012fr}. 
In fact, in \cite{Csaki:2011ge} they also get a hierarchy where the third generation couplings
dominate with fixed ratios between them.
Fixing the expansion parameter as $\theta=0.22$, their set of
$R$-parity breaking parameters can be written as
\begin{align}
\label{eq:Grosslpp}
 \begin{pmatrix}
    \lambda_{112}''&\lambda_{212}''&\lambda_{312}''\\
    \lambda_{113}''&\lambda_{213}''&\lambda_{313}''\\
    \lambda_{123}''&\lambda_{223}''&\lambda_{323}''\\
  \end{pmatrix}\approx&
 \tan^2\beta_{\text{MFV}}\begin{pmatrix}
   \theta^{24} & \theta^{18} & \theta^{13}\\
   \theta^{19} & \theta^{14} & \theta^{12}\\
   \theta^{16} & \theta^{13} & \theta^{11}\\
  \end{pmatrix}=
 \theta^{n_{\text{MFV}}}\begin{pmatrix}
   \theta^{13} & \theta^{7} & \theta^2\\
   \theta^8 & \theta^3 & \theta\\
   \theta^5 & \theta^2 & 1\\
  \end{pmatrix}\,,
\end{align}
with $\theta^{n_{\text{MFV}}}=\theta^{11}\tan^2\beta_{\text{MFV}}$. 
Comparing with eq.~\eqref{eq:Hierlpp}, we can see that the set of
predicted couplings until order $\theta^{n_{\text{MFV}}+3}$, is
basically the same than in our case (with the exception of their
$\lambda_{312}''$ which has an additional suppression factor of
$\theta$).
Therefore, the phenomenology of both theories for $R$-parity violation
should be the same at LHC. 
In fact, the phenomenology of~\cite{Csaki:2011ge} for the leading
couplings, was analyzed in detail at the  LHC with the results
presented as function of $\tan\beta_{\text{MFV}}$.
The specific values at
$\tan\beta_{\text{MFV}}\approx(44.5,20.7,9.7,4.6,2.1)$ in the several
plots of~\cite{Csaki:2011ge} corresponds to the discrete set of solutions
$n_{\lambda''}=(6,7,8,9,10)$ respectively, in our model. 
In particular several plots there, they explore the decay length ($c\tau$)
for LSP masses in the range $100-800\ \text{GeV}$.
When the stop is the LSP for example, displaced vertices (DV) are
expected for $n_{\lambda''}=10$.
For a sbottom LSP it is possible to have DV for $n_{\lambda''}=9,10$,
while the three-body decays of a LSP neutralino could generate DV for
$n_{\lambda''}=8,9,10$.
In the same vein, because decays of the stau LSP involves four
particles in the final state, DV are expected for $n_{\lambda''}\ge
6$.

Recent phenomenological analysis in $R$-parity breaking
trough $UDD$ operators have focused in prompt decay for stops and
sbottoms~\cite{Csaki:2011ge,Franceschini:2012za,Bhattacherjee:2013gr,Csaki:2013we}.
However, the experimental results about DV at LHC
are in general not directly applicable to this kind of models, because
high $p_T$ leptons are required for trigger the
events~\cite{Aad:2011zb,Aad:2012zx,Chatrchyan:2012jna}, and to be part of the
DV~\cite{Aad:2012zx,Chatrchyan:2012jna}. 
We assume in the discussion below that pure hadronic DV are still
compatible with light squarks and gluinos.

Regarding collider searches, a pair produced gluino with a prompt decay to three
jets have been searched by CDF \cite{Aaltonen:2011sg}, CMS
\cite{Chatrchyan:2011cj,Chatrchyan:2012uxa} and ATLAS \cite{ATLAS:2012dp}.%
\footnote{In this analysis all the superpartners except for the
  gluinos are decoupled, and some reinterpretation would be needed to
  apply the results to a more generic SUSY spectrum.}
CMS results constrain the gluino mass to be in the ranges
$144<m_{\tilde{g}}<200$ GeV or $m_{\tilde{g}}>460$ GeV.
However, ATLAS already excludes gluino masses up to
$m_{\tilde{g}}\lesssim666$ GeV.
In general, this bounds do not apply when the gluino is not the
LSP~\cite{Csaki:2011ge,Bhattacherjee:2013gr}.  
On the other hand, CDF \cite{Aaltonen:2008dn}, ATLAS
\cite{Aad:2011yh,ATLAS:2012ds} and CMS \cite{Chatrchyan:2013izb} also
have performed searches for pair production of dijet resonances in
four-jet events without putting appreciable constraints on stops
decaying to dijets.
Therefore, the already analyzed data at LHC still allows for low
squarks and gluinos in scenarios with $R$-parity breaking through
$B$-violating couplings~\cite{Evans:2012bf,Bhattacherjee:2013gr} 

We have seen that both this single-flavon horizontal (SFH) and the MFV models, lead to a
realistic and predictive framework which could be more easily probed
at LHC than some {\it ad hoc} version or $R$-parity breaking with $B$
violation.  
In fact, recently in~\cite{Berger:2013sir} the CMS results on searches for
new physics in events with same-sign dileptons and $b$ jets
\cite{Chatrchyan:2012paa} have been recasting in a simplified version of the
$R$-parity breaking MFV model where it is assumed one spectrum with
only two light states: a gluino and a stop. 
All other SUSY particles are assumed to be either too heavy or too
weakly coupled to be relevant at the LHC. 
Furthermore, the stop is assumed to be the LSP, and
$m_{\tilde{g}}>m_{\tilde{t}}+m_t$.\footnote{As a consequence the gluino branching to stop-top is equal to 1.}  
Under these conditions they are able to set a lower bound on the gluino
mass about $800\ \text{GeV}$ at 95\% of confidence level.\footnote{The
  obtained lower bound only apply if the gluino is a Majorana
  particle.}
The same bound could apply to the SFH model with
$R$-parity breaking presented in this work.

In order to really probe this single-flavon horizontal (or the MFV) 
$R$-parity breaking model, the full textures in eq.~\eqref{eq:Hlpp} or \eqref{eq:Grosslpp}
should be probed.
However, relations between different branching ratios could be
measured only in $e^+e^-$ colliders.
In a stop LSP scenario, it can decay directly into two down quarks of
different generations through the $\lambda''_{3jk}$ coupling.  In this
case, the hierarchy between $\lambda''$ couplings allows for estimate
several fractions of branchings, e.g.  $\operatorname{Br}(\tilde t\to
\bar{s}\bar{b})/\operatorname{Br}(\tilde t\to
\bar{d}\bar{s})/\sim\theta^2$.
A sbottom LSP, with a mass larger than the top mass, may show the
clear hierarchy $\operatorname{Br}(\tilde b\to
\bar{t}\bar{s})/\operatorname{Br}(\tilde b\to
\bar{c}\bar{s})/\sim\theta^4$.
For a neutralino LSP with $m_{\tilde\chi}^0>m_t$, the dominant
coupling $\lambda''_{323}$ entails $\operatorname{Br}(\tilde\chi\to t
d b)/\operatorname{Br}(\tilde\chi\to t s
b)\sim\operatorname{Br}(\tilde\chi\to t d
s)/\operatorname{Br}(\tilde\chi\to t s b)\sim\theta^2$ and
$\operatorname{Br}(\tilde\chi\to c s
b)/\operatorname{Br}(\tilde\chi\to t s b)\sim\theta^4$.
For the case $m_{\tilde\chi}<m_t$ the main neutralino decay is then
controlled by $\lambda''_{223}$, and  will produce charm quarks with
ratios of branching ratios given by $\operatorname{Br}(\tilde\chi\to c
d b)/\operatorname{Br}(\tilde\chi\to c s
b)\sim\operatorname{Br}(\tilde\chi\to c d
s)/\operatorname{Br}(\tilde\chi\to c s b)\sim\theta^2$.


\section{Conclusions}\label{sec:conlusions}
We have obtained a supersymmetric $R$-parity breaking model with $B$
violation, by considering the most general supersymmetric standard
model allowed by Gauge invariance, and extending it with a
single-flavon horizontal (SFH) $U(1)_H$ symmetry. 
The generated effective theory at low energy, have only the particle
content of the SSM.
After imposing existing constraints in both single and quadratic
R-parity violating (RPV) couplings, only one precise
hierarchy remains depending in a global suppression factor
$\theta^{n_{\lambda''}}$ ($n_{\lambda}''>1$) with $\lambda''_{323}$ as
the dominant coupling, and very suppressed couplings for the first two
generations.
Additional suppression is required in order to obtain Dirac neutrino
masses in the model, and only solutions with $n_{\lambda}''\ge6$
remain allowed.    
In this way, the resulting RPV and $B$ violating model also explaining
neutrino masses, is powerful enough to satisfy all the existing constraints
on RPV. 
In particular, the $U(1)_H$ symmetry also ensures that dimension-5
$L$ violating operators are sufficiently suppressed so that
the decay of the proton is above the experimental limits.

The resulting underlying theory for the RPV operators, is quite
similar to the obtained after imposing the Minimal Flavor Violation
(MFV) hypothesis on a general RPV model (at least until couplings of
order $\theta^{n_{\lambda''}+3}$) and therefore the predictions of
both models are the same at the LHC.

The phenomenology at colliders depends strongly on the nature and decay length of
the LSP. 
Specific searches at LHC for the RPV with $B$-violation have reported
restrictions only in the case of prompt decays of the gravitino when
it is the LSP.  
Several analysis of CMS and ATLAS involving leptons, have been
reanalyzed to constraint the gluino as function of the stop mass
(\cite{Berger:2013sir} and references therein) within a special spectrum
guaranteeing that $\operatorname{BR}(\tilde{g}\to
\tilde{t}\,\bar{t})=1$, and with prompt decays of the corresponding
LSP stop. 
In both cases bounds in the gluino mass around $600\text{ GeV}$ have
been obtained. Therefore, the parameter space of the RPV/SFH scenario
(or the RPV/MFV one) have still plenty of room to accommodate a low
energy supersymmetric spectrum.

There are a number of open issues that could be more easily studied
within this realistic and predictive framework.  For example: the
constraints on the couplings from low energy observables, and indirect
dark matter experiments; or the restrictions in the parameter space
from other collider signatures like the displaced vertices searches
already implemented by ATLAS~\cite{Aad:2012zx} and
CMS~\cite{Chatrchyan:2012jna}.


\section{Acknowledgments}
DR and OZ have been partially supported by COLCIENCIAS through the grant number 111-556-934918, UdeA/CODI grant IN624CE and Sostenibilidad-UdeA. 

\appendix{}

\section{Integer part of $R$-parity breaking $H$-charges}
\label{sec:integer-part-r}
Functions of the trilinear $R$-parity breaking coupling indices
returning \emph{integer} values:
\begin{align}
  \label{eq:Is}
  \mathcal{I}''(ijk)=&-2i+p_i''+p_j''+p_k''&(j<&k)\nonumber\\
  \mathcal{I}'(ijk)=&\frac{1}{2}\left(j+k+p_j'+p_k'\right)-2\delta_{j3}&&     \nonumber\\
  \mathcal{I}(ijk)=&i-2k+p_i+p_j+p_k\,,        &(i<&j)
\end{align}
where the several $p_i$'s are shown in Table~\ref{tab:pi}.
\begin{table}
  \centering
  \begin{tabular}{llll}\hline
    $i$ & $1$ & $2$ & $3$\\\hline
    $p_i$ & $3$ & $2$ & $2$\\
    $p_i'$ & $4$ & $1$ & $0$\\
    $p_i''$ & $3$ & $2$ & $2$\\\hline
  \end{tabular}
  \caption{Integer values requiered to obtain the horizontal charges of dimension-4 RPV operators.}
  \label{tab:pi}
\end{table}

\section{$H$-charges of dimension-5 operators}\label{app:D5}
The horizontal charges for the dimension-5 operators that violate only $B$ are given by
\begin{align}\label{eq:D5BNV1}\nonumber
H\left[(\kappa_3)_{1jk}\widehat{Q}_1\widehat{Q}_j\widehat{Q}_k\widehat{H}_d\right]=&
A_3+(2x+4-n_{\lambda''})\mathbf{1_3},\\\nonumber
H\left[(\kappa_3)_{2jk}\widehat{Q}_2\widehat{Q}_j\widehat{Q}_k\widehat{H}_d\right]=&
A_3+(2x+3-n_{\lambda''})\mathbf{1_3},\\
H\left[(\kappa_3)_{3jk}\widehat{Q}_3\widehat{Q}_j\widehat{Q}_k\widehat{H}_d\right]=&
A_3+(2x+1-n_{\lambda''})\mathbf{1_3},
\end{align}
\begin{align}\label{eq:D5BNV2}\nonumber
H\left[(\kappa_{10})_{ij1}\widehat{Q}_i\widehat{Q}_j\widehat{d}^*_1\right]=&
A_3+(x-n_{\lambda''})\mathbf{1_3},\\\nonumber
H\left[(\kappa_{10})_{ij2}\widehat{Q}_i\widehat{Q}_j\widehat{d}^*_2\right]=&
A_3+(x+1-n_{\lambda''})\mathbf{1_3},\\
H\left[(\kappa_{10})_{ij2}\widehat{Q}_i\widehat{Q}_j\widehat{d}^*_3\right]=&H\left[(\kappa_{10})_{ij3}\widehat{Q}_i\widehat{Q}_j\widehat{d}^*_2\right].
\end{align}
For the lepton and baryon-number violating operators we have that
\begin{align}\label{eq:D5BLNV1}\nonumber
 H\left[(\kappa_{1})_{1jkl}\widehat{Q}_1\widehat{Q}_j\widehat{Q}_k\widehat{L}_l\right]=&
A_1+(5+2x+n_l-n_{\lambda''})\mathbf{1_3},\\\nonumber
 H\left[(\kappa_{1})_{2jkl}\widehat{Q}_2\widehat{Q}_j\widehat{Q}_k\widehat{L}_l\right]=&
A_1+(4+2x+n_l-n_{\lambda''})\mathbf{1_3},\\
 H\left[(\kappa_{1})_{3jkl}\widehat{Q}_3\widehat{Q}_j\widehat{Q}_k\widehat{L}_l\right]=&
A_1+(2+2x+n_l-n_{\lambda''})\mathbf{1_3},
\end{align}
\begin{align}\label{eq:D5BLNV2}\nonumber
H\left[(\kappa_{2})_{ij11}\widehat{u}_i\widehat{u}_j\widehat{d}_1\widehat{e}_1\right]=&
A_2+(6-n_1+n_{\lambda''})\mathbf{1_3},\\\nonumber
H\left[(\kappa_{2})_{ij21}\widehat{u}_i\widehat{u}_j\widehat{d}_2\widehat{e}_1\right]=&
A_2+(5-n_1+n_{\lambda''})\mathbf{1_3},\\\nonumber
H\left[(\kappa_{2})_{ij31}\widehat{u}_i\widehat{u}_j\widehat{d}_3\widehat{e}_1\right]=&
H\left[(\kappa_{2})_{ij21}\widehat{u}_i\widehat{u}_j\widehat{d}_2\widehat{e}_1\right],\\\nonumber
H\left[(\kappa_{2})_{ij12}\widehat{u}_i\widehat{u}_j\widehat{d}_1\widehat{e}_2\right]=&
A_2+(3-n_2+n_{\lambda''})\mathbf{1_3},\\\nonumber
H\left[(\kappa_{2})_{ij22}\widehat{u}_i\widehat{u}_j\widehat{d}_2\widehat{e}_2\right]=&
A_2+(2-n_2+n_{\lambda''})\mathbf{1_3},\\\nonumber
H\left[(\kappa_{2})_{ij32}\widehat{u}_i\widehat{u}_j\widehat{d}_2\widehat{e}_2\right]=&H\left[(\kappa_{2})_{ij22}\widehat{u}_i\widehat{u}_j\widehat{d}_3\widehat{e}_2\right]\\\nonumber
H\left[(\kappa_{2})_{ij13}\widehat{u}_i\widehat{u}_j\widehat{d}_1\widehat{e}_3\right]=&
A_2+(1-n_3+n_{\lambda''})\mathbf{1_3},\\\nonumber
H\left[(\kappa_{2})_{ij23}\widehat{u}_i\widehat{u}_j\widehat{d}_2\widehat{e}_3\right]=&
A_2+(-n_3+n_{\lambda''})\mathbf{1_3},\\
H\left[(\kappa_{2})_{ij33}\widehat{u}_i\widehat{u}_j\widehat{d}_2\widehat{e}_3\right]=&H\left[(\kappa_{2})_{ij23}\widehat{u}_i\widehat{u}_j\widehat{d}_3\widehat{e}_3\right].
\end{align}
Finally, for the lepton-number violating terms we have found
\begin{align}\label{eq:D5LNV}\nonumber
H\left[(\kappa_{4})_{ij1}\widehat{Q}_i\widehat{H}_d\widehat{u}_j\widehat{e}_1\right]=&
A_4+(5-n_1+x)\mathbf{1_3},\\\nonumber
H\left[(\kappa_{4})_{ij2}\widehat{Q}_i\widehat{H}_d\widehat{u}_j\widehat{e}_2\right]=&
A_4+(2-n_2+x)\mathbf{1_3},\\\nonumber
H\left[(\kappa_{4})_{ij3}\widehat{Q}_i\widehat{H}_d\widehat{u}_j\widehat{e}_3\right]=&
A_4+(-n_3+x)\mathbf{1_3},\\\nonumber
H\left[(\kappa_{5})_{ij}\widehat{L}_i\widehat{H}_u\widehat{L}_j\widehat{H}_u\right]=&
\left(
\begin{array}{ccc}
 2n_1 & n_1+n_2 & n_1+n_3 \\
 n_1+n_2 & 2n_2 & n_2+n_3 \\
 n_1+n_3 & n_2+n_3 & 2n_3
\end{array}
\right),\\\nonumber
H\left[(\kappa_{6})_{i}\widehat{L}_i\widehat{H}_u\widehat{H}_d\widehat{H}_u\right]=&-1+n_i,\\\nonumber
H\left[(\kappa_{7})_{ij1}\widehat{u}_i\widehat{d}^*_j\widehat{e}_1\right]=&
A_7+(4-n_1)\mathbf{1_3},\\\nonumber
H\left[(\kappa_{7})_{ij2}\widehat{u}_i\widehat{d}^*_j\widehat{e}_2\right]=&
A_7+(1-n_2)\mathbf{1_3},\\\nonumber
H\left[(\kappa_{7})_{ij3}\widehat{u}_i\widehat{d}^*_j\widehat{e}_3\right]=&
A_7+(-1-n_3)\mathbf{1_3},\\\nonumber
H\left[(\kappa_{8})_{1}\widehat{H}_u^*\widehat{H}_d\widehat{e}_1\right]=&5-n_1+x,\\\nonumber
H\left[(\kappa_{8})_{2}\widehat{H}_u^*\widehat{H}_d\widehat{e}_2\right]=&2-n_2+x,\\\nonumber
H\left[(\kappa_{8})_{3}\widehat{H}_u^*\widehat{H}_d\widehat{e}_3\right]=&-n_3+x,\\
H\left[(\kappa_{9})_{i1k}\widehat{Q}_i\widehat{L}^*_j\widehat{u}_k\right]=&
A_9+(-n_j)\mathbf{1_3}.
\end{align}
In the above expressions we have defined 
\begin{align}
A_1&=A_3=\left(
\begin{array}{ccc}
 6 & 5 & 3 \\
 5 & 4 & 2 \\
 3 & 2 & 0
\end{array}
\right), \, 
A_2=\left(
\begin{array}{ccc}
 10 & 7 & 5 \\
 7 & 4 & 2 \\
 5 & 2 & 0
\end{array}
\right),\,
A_4=A_9=\left(
\begin{array}{ccc}
 8 & 5 & 3 \\
 7 & 4 & 2 \\
 5 & 2 & 0
\end{array}
\right),\,
A_7=\left(
\begin{array}{ccc}
 5 & 6 & 6 \\
 2 & 3 & 3 \\
 0 & 1 & 1
\end{array}
\right).
\end{align}


\bibliographystyle{h-physrev4}
\bibliography{susy}

\begin{thebibliography}{10}

\bibitem{:2012mfa}
CMS Collaboration, S.~Chatrchyan {\em et~al.},
\newblock (2012), arXiv:1207.1898.

\bibitem{:2012rz}
ATLAS Collaboration, G.~Aad {\em et~al.},
\newblock (2012), arXiv:1208.0949.

\bibitem{Hall:1983id}
L.~J. Hall and M.~Suzuki,
\newblock Nucl. Phys. {\bf B231}, 419 (1984).

\bibitem{Ross:1984yg}
G.~G. Ross and J.~Valle,
\newblock Phys.Lett. {\bf B151}, 375 (1985).

\bibitem{Barger:1989rk}
V.~D. Barger, G.~Giudice, and T.~Han,
\newblock Phys.Rev. {\bf D40}, 2987 (1989).

\bibitem{Dreiner:1997uz}
H.~K. Dreiner,
\newblock (1997), arXiv:hep-ph/9707435.

\bibitem{Allanach:2003eb}
B.~C. Allanach, A.~Dedes, and H.~K. Dreiner,
\newblock Phys. Rev. {\bf D69}, 115002 (2004), arXiv:hep-ph/0309196.

\bibitem{Barbier:2004ez}
R.~Barbier {\em et~al.},
\newblock Phys. Rept. {\bf 420}, 1 (2005), arXiv:hep-ph/0406039.

\bibitem{Takayama:2000uz}
F.~Takayama and M.~Yamaguchi,
\newblock Phys. Lett. {\bf B485}, 388 (2000), arXiv:hep-ph/0005214.

\bibitem{Buchmuller:2007ui}
W.~Buchmuller, L.~Covi, K.~Hamaguchi, A.~Ibarra, and T.~Yanagida,
\newblock JHEP {\bf 03}, 037 (2007), arXiv:hep-ph/0702184.

\bibitem{deCampos:2007bn}
F.~de~Campos {\em et~al.},
\newblock JHEP {\bf 05}, 048 (2008), arXiv:0712.2156.

\bibitem{Brust:2011tb}
C.~Brust, A.~Katz, S.~Lawrence, and R.~Sundrum,
\newblock JHEP {\bf 1203}, 103 (2012), arXiv:1110.6670.

\bibitem{Butterworth:2009qa}
J.~M. Butterworth, J.~R. Ellis, A.~R. Raklev, and G.~P. Salam,
\newblock Phys. Rev. Lett. {\bf 103}, 241803 (2009), arXiv:0906.0728.

\bibitem{Allanach:2012vj}
B.~Allanach and B.~Gripaios,
\newblock JHEP {\bf 1205}, 062 (2012), arXiv:1202.6616.

\bibitem{Brust:2012uf}
C.~Brust, A.~Katz, and R.~Sundrum,
\newblock JHEP {\bf 1208}, 059 (2012), arXiv:1206.2353.

\bibitem{Asano:2012gj}
M.~Asano, K.~Rolbiecki, and K.~Sakurai,
\newblock (2012), arXiv:1209.5778.

\bibitem{Curtin:2012rm}
D.~Curtin, R.~Essig, and B.~Shuve,
\newblock (2012), arXiv:1210.5523.

\bibitem{Graham:2012th}
P.~W. Graham, D.~E. Kaplan, S.~Rajendran, and P.~Saraswat,
\newblock JHEP {\bf 1207}, 149 (2012), arXiv:1204.6038.

\bibitem{Han:2012cu}
Z.~Han, A.~Katz, M.~Son, and B.~Tweedie,
\newblock (2012), arXiv:1211.4025.

\bibitem{Ibanez:1991pr}
L.~E. Ibanez and G.~G. Ross,
\newblock Nucl.Phys. {\bf B368}, 3 (1992).

\bibitem{Dreiner:2005rd}
H.~K. Dreiner, C.~Luhn, and M.~Thormeier,
\newblock Phys.Rev. {\bf D73}, 075007 (2006), arXiv:hep-ph/0512163.

\bibitem{Tamvakis:1996np}
K.~Tamvakis,
\newblock Phys.Lett. {\bf B382}, 251 (1996), arXiv:hep-ph/9604343.

\bibitem{Eyal:1999gq}
G.~Eyal and Y.~Nir,
\newblock JHEP {\bf 9906}, 024 (1999), arXiv:hep-ph/9904473.

\bibitem{Nikolidakis:2007fc}
E.~Nikolidakis and C.~Smith,
\newblock Phys.Rev. {\bf D77}, 015021 (2008), arXiv:0710.3129.

\bibitem{Luhn:2007gq}
C.~Luhn and M.~Thormeier,
\newblock Phys.Rev. {\bf D77}, 056002 (2008), arXiv:0711.0756.

\bibitem{Csaki:2011ge}
C.~Csaki, Y.~Grossman, and B.~Heidenreich,
\newblock Phys.Rev. {\bf D85}, 095009 (2012), arXiv:1111.1239.

\bibitem{KerenZur:2012fr}
B.~Keren-Zur {\em et~al.},
\newblock Nucl.Phys. {\bf B867}, 429 (2013), arXiv:1205.5803.

\bibitem{FileviezPerez:2011pt}
P.~Fileviez~Perez and M.~B. Wise,
\newblock JHEP {\bf 1108}, 068 (2011), arXiv:1106.0343.

\bibitem{Ruderman:2012jd}
J.~T. Ruderman, T.~R. Slatyer, and N.~Weiner,
\newblock (2012), arXiv:1207.5787.

\bibitem{Dreiner:2012ae}
H.~K. Dreiner, M.~Hanussek, and C.~Luhn,
\newblock Phys.Rev. {\bf D86}, 055012 (2012), arXiv:1206.6305.

\bibitem{Krnjaic:2012aj}
G.~Krnjaic and D.~Stolarski,
\newblock (2012), arXiv:1212.4860.

\bibitem{Csaki:2013we}
C.~Csaki and B.~Heidenreich,
\newblock (2013), arXiv:1302.0004.

\bibitem{Franceschini:2013ne}
R.~Franceschini and R.~Mohapatra,
\newblock (2013), arXiv:1301.3637.

\bibitem{Froggatt:1978nt}
C.~D. Froggatt and H.~B. Nielsen,
\newblock Nucl. Phys. {\bf B147}, 277 (1979).

\bibitem{Sierra:2009zq}
D.~Aristizabal~Sierra, D.~Restrepo, and O.~Zapata,
\newblock Phys. Rev. {\bf D80}, 055010 (2009), arXiv:0907.0682.

\bibitem{Dreiner:2003hw}
H.~K. Dreiner and M.~Thormeier,
\newblock Phys. Rev. {\bf D69}, 053002 (2004), arXiv:hep-ph/0305270.

\bibitem{Choi:1998wc}
K.~Choi, K.~Hwang, and E.~J. Chun,
\newblock Phys. Rev. {\bf D60}, 031301 (1999), arXiv:hep-ph/9811363.

\bibitem{Mira:2000gg}
J.~M. Mira, E.~Nardi, D.~A. Restrepo, and J.~W.~F. Valle,
\newblock Phys. Lett. {\bf B492}, 81 (2000), arXiv:hep-ph/0007266.

\bibitem{Dreiner:2003yr}
H.~K. Dreiner, H.~Murayama, and M.~Thormeier,
\newblock Nucl. Phys. {\bf B729}, 278 (2005), arXiv:hep-ph/0312012.

\bibitem{Dreiner:2006xw}
H.~K. Dreiner, C.~Luhn, H.~Murayama, and M.~Thormeier,
\newblock Nucl. Phys. {\bf B774}, 127 (2007), arXiv:hep-ph/0610026.

\bibitem{Dreiner:2007vp}
H.~K. Dreiner, C.~Luhn, H.~Murayama, and M.~Thormeier,
\newblock Nucl. Phys. {\bf B795}, 172 (2008), arXiv:0708.0989.

\bibitem{Hall:1999sn}
L.~J. Hall, H.~Murayama, and N.~Weiner,
\newblock Phys.Rev.Lett. {\bf 84}, 2572 (2000), arXiv:hep-ph/9911341.

\bibitem{Haba:2000be}
N.~Haba and H.~Murayama,
\newblock Phys.Rev. {\bf D63}, 053010 (2001), arXiv:hep-ph/0009174.

\bibitem{deGouvea:2003xe}
A.~de~Gouvea and H.~Murayama,
\newblock Phys.Lett. {\bf B573}, 94 (2003), arXiv:hep-ph/0301050.

\bibitem{deGouvea:2012ac}
A.~de~Gouvea and H.~Murayama,
\newblock (2012), arXiv:1204.1249.

\bibitem{Abe:2011sj}
T2K Collaboration, K.~Abe {\em et~al.},
\newblock Phys.Rev.Lett. {\bf 107}, 041801 (2011), arXiv:1106.2822.

\bibitem{Abe:2011fz}
DOUBLE-CHOOZ Collaboration, Y.~Abe {\em et~al.},
\newblock Phys.Rev.Lett. {\bf 108}, 131801 (2012), arXiv:1112.6353.

\bibitem{An:2012eh}
DAYA-BAY Collaboration, F.~An {\em et~al.},
\newblock Phys.Rev.Lett. {\bf 108}, 171803 (2012), arXiv:1203.1669.

\bibitem{Ahn:2012nd}
RENO collaboration, J.~Ahn {\em et~al.},
\newblock Phys.Rev.Lett. {\bf 108}, 191802 (2012), arXiv:1204.0626.

\bibitem{Irges:1998ax}
N.~Irges, S.~Lavignac, and P.~Ramond,
\newblock Phys. Rev. {\bf D58}, 035003 (1998), arXiv:hep-ph/9802334.

\bibitem{Choi:1996se}
K.~Choi, E.~J. Chun, and H.~D. Kim,
\newblock Phys. Lett. {\bf B394}, 89 (1997), arXiv:hep-ph/9611293.

\bibitem{Joshipura:2000sn}
A.~S. Joshipura, R.~D. Vaidya, and S.~K. Vempati,
\newblock Phys. Rev. {\bf D62}, 093020 (2000), arXiv:hep-ph/0006138.

\bibitem{Binetruy:1996xk}
P.~Binetruy, S.~Lavignac, and P.~Ramond,
\newblock Nucl. Phys. {\bf B477}, 353 (1996), arXiv:hep-ph/9601243.

\bibitem{Ellis:1998rj}
J.~R. Ellis, S.~Lola, and G.~G. Ross,
\newblock Nucl.Phys. {\bf B526}, 115 (1998), arXiv:hep-ph/9803308.

\bibitem{BenHamo:1994bq}
V.~Ben-Hamo and Y.~Nir,
\newblock Phys.Lett. {\bf B339}, 77 (1994), arXiv:hep-ph/9408315.

\bibitem{Dudas:1995yu}
E.~Dudas, S.~Pokorski, and C.~A. Savoy,
\newblock Phys. Lett. {\bf B356}, 45 (1995), arXiv:hep-ph/9504292.

\bibitem{Green:1984sg}
M.~B. Green and J.~H. Schwarz,
\newblock Phys. Lett. {\bf B149}, 117 (1984).

\bibitem{Diaz:2003as}
M.~A. Diaz, M.~Hirsch, W.~Porod, J.~C. Romao, and J.~W.~F. Valle,
\newblock Phys. Rev. {\bf D68}, 013009 (2003), arXiv:hep-ph/0302021.

\bibitem{Goity:1994dq}
J.~Goity and M.~Sher,
\newblock Phys.Lett. {\bf B346}, 69 (1995), arXiv:hep-ph/9412208.

\bibitem{Berger:1991fa}
Frejus Collaboration, C.~Berger {\em et~al.},
\newblock Phys.Lett. {\bf B269}, 227 (1991).

\bibitem{Litos:2010}
M.~D. Litos,
\newblock PhD thesis, Boston University  (2010).

\bibitem{BarShalom:2002sv}
S.~Bar-Shalom, G.~Eilam, and Y.-D. Yang,
\newblock Phys.Rev. {\bf D67}, 014007 (2003), arXiv:hep-ph/0201244.

\bibitem{Aubert:2006nn}
BABAR Collaboration, B.~Aubert {\em et~al.},
\newblock Phys.Rev. {\bf D74}, 011102 (2006), arXiv:hep-ex/0605037.

\bibitem{Kim:2012gt}
Belle Collaboration, J.~Kim {\em et~al.},
\newblock Phys.Rev. {\bf D86}, 031101 (2012), arXiv:1206.4760.

\bibitem{Giudice:1988yz}
G.~F. Giudice and A.~Masiero,
\newblock Phys. Lett. {\bf B206}, 480 (1988).

\bibitem{Choi:1996nk}
K.~Choi, E.~J. Chun, and J.~S. Lee,
\newblock Phys.Rev. {\bf D55}, 3924 (1997), arXiv:hep-ph/9611285.

\bibitem{Lola:2008bk}
N.-E. Bomark, S.~Lola, P.~Osland, and A.~Raklev,
\newblock Phys.Lett. {\bf B677}, 62 (2009), arXiv:0811.2969.

\bibitem{Lola:2007rw}
S.~Lola, P.~Osland, and A.~Raklev,
\newblock Phys.Lett. {\bf B656}, 83 (2007), arXiv:0707.2510.

\bibitem{Ibanez:1991hv}
L.~E. Ibanez and G.~G. Ross,
\newblock Phys.Lett. {\bf B260}, 291 (1991).

\bibitem{Altarelli:2002sg}
G.~Altarelli, F.~Feruglio, and I.~Masina,
\newblock JHEP {\bf 0301}, 035 (2003), arXiv:hep-ph/0210342.

\bibitem{Buchmuller:2011tm}
W.~Buchmuller, V.~Domcke, and K.~Schmitz,
\newblock JHEP {\bf 1203}, 008 (2012), arXiv:1111.3872.

\bibitem{Altarelli:2012ia}
G.~Altarelli, F.~Feruglio, I.~Masina, and L.~Merlo,
\newblock (2012), arXiv:1207.0587.

\bibitem{Jack:2003pb}
I.~Jack, D.~Jones, and R.~Wild,
\newblock Phys.Lett. {\bf B580}, 72 (2004), arXiv:hep-ph/0309165.

\bibitem{Dreiner:2008ca}
H.~K. Dreiner and S.~Grab,
\newblock Phys. Lett. {\bf B679}, 45 (2009), arXiv:0811.0200.

\bibitem{Evans:2012bf}
J.~A. Evans and Y.~Kats,
\newblock (2012), arXiv:1209.0764.

\bibitem{Berger:2012mm}
J.~Berger, C.~Csaki, Y.~Grossman, and B.~Heidenreich,
\newblock (2012), arXiv:1209.4645.

\bibitem{Carpenter:2008sy}
L.~M. Carpenter, D.~E. Kaplan, and E.~J. Rhee,
\newblock (2008), arXiv:0804.1581.

\bibitem{Dreiner:2012wm}
H.~Dreiner, F.~Staub, A.~Vicente, and W.~Porod,
\newblock Phys.Rev. {\bf D86}, 035021 (2012), arXiv:1205.0557.

\bibitem{Franceschini:2012za}
R.~Franceschini and R.~Torre,
\newblock (2012), arXiv:1212.3622.

\bibitem{Bomark:2011ye}
N.-E. Bomark, D.~Choudhury, S.~Lola, and P.~Osland,
\newblock JHEP {\bf 1107}, 070 (2011), arXiv:1105.4022.

\bibitem{Choudhury:2011ve}
D.~Choudhury, M.~Datta, and M.~Maity,
\newblock JHEP {\bf 1110}, 004 (2011), arXiv:1106.5114.

\bibitem{Desai:2010sq}
N.~Desai and B.~Mukhopadhyaya,
\newblock JHEP {\bf 1010}, 060 (2010), arXiv:1002.2339.

\bibitem{Kilic:2011sr}
C.~Kilic and S.~Thomas,
\newblock Phys.Rev. {\bf D84}, 055012 (2011), arXiv:1104.1002.

\bibitem{Carpenter:2007zz}
L.~M. Carpenter, D.~E. Kaplan, and E.-J. Rhee,
\newblock Phys.Rev.Lett. {\bf 99}, 211801 (2007), arXiv:hep-ph/0607204.

\bibitem{Kaplan:2007ap}
D.~E. Kaplan and K.~Rehermann,
\newblock JHEP {\bf 10}, 056 (2007), arXiv:0705.3426.

\bibitem{Berger:2013sir}
J.~Berger, M.~Perelstein, M.~Saelim, and P.~Tanedo,
\newblock (2013), arXiv:1302.2146.

\bibitem{Bhattacherjee:2013gr}
B.~Bhattacherjee, J.~L. Evans, M.~Ibe, S.~Matsumoto, and T.~T. Yanagida,
\newblock (2013), arXiv:1301.2336.

\bibitem{Aad:2011zb}
ATLAS Collaboration, G.~Aad {\em et~al.},
\newblock Phys.Lett. {\bf B707}, 478 (2012), arXiv:1109.2242.

\bibitem{Aad:2012zx}
ATLAS Collaboration, G.~Aad {\em et~al.},
\newblock Phys.Lett. {\bf B719}, 280 (2013), arXiv:1210.7451.

\bibitem{Chatrchyan:2012jna}
CMS Collaboration, S.~Chatrchyan {\em et~al.},
\newblock JHEP {\bf 1302}, 085 (2013), arXiv:1211.2472.

\bibitem{Aaltonen:2011sg}
CDF Collaboration, T.~Aaltonen {\em et~al.},
\newblock Phys.Rev.Lett. {\bf 107}, 042001 (2011), arXiv:1105.2815.

\bibitem{Chatrchyan:2011cj}
CMS Collaboration, S.~Chatrchyan {\em et~al.},
\newblock Phys.Rev.Lett. {\bf 107}, 101801 (2011), arXiv:1107.3084.

\bibitem{Chatrchyan:2012uxa}
CMS Collaboration, S.~Chatrchyan {\em et~al.},
\newblock Phys.Lett. {\bf B718}, 329 (2012), arXiv:1208.2931.

\bibitem{ATLAS:2012dp}
ATLAS Collaboration, G.~Aad {\em et~al.},
\newblock (2012), arXiv:1210.4813.

\bibitem{Aaltonen:2008dn}
CDF Collaboration, T.~Aaltonen {\em et~al.},
\newblock Phys.Rev. {\bf D79}, 112002 (2009), arXiv:0812.4036.

\bibitem{Aad:2011yh}
ATLAS Collaboration, G.~Aad {\em et~al.},
\newblock Eur.Phys.J. {\bf C71}, 1828 (2011), arXiv:1110.2693.

\bibitem{ATLAS:2012ds}
ATLAS Collaboration, G.~Aad {\em et~al.},
\newblock (2012), arXiv:1210.4826.

\bibitem{Chatrchyan:2013izb}
CMS Collaboration, S.~Chatrchyan {\em et~al.},
\newblock (2013), arXiv:1302.0531.

\bibitem{Chatrchyan:2012paa}
CMS Collaboration, S.~Chatrchyan {\em et~al.},
\newblock (2012), arXiv:1212.6194.

\end{thebibliography}

\end{document}